\documentclass[iop]{emulateapj}
\usepackage{natbib}
\usepackage{longtable}
\usepackage{amsmath}
\usepackage{color}
\def\dt{\Delta t}
\def\taud{\tau_\mathrm{d}}
\def\spear{{\tt SPEAR}}
\def\javelin{{\tt JAVELIN}}
\def\gp{\mathcal{GP}}
\def\civ{C\,{\sc iv}}
\def\mg2{Mg\,{\sc ii}}
\def\ha{H$\alpha$}
\def\hb{H$\beta$}

\def\bfy{{\bf y}}
\def\bfs{{\bf s}}
\def\bfhq{{\bf\hat{q}}}
\def\bfhs{{\bf\hat{s}}}
\def\cb{\mathbb{C}}
\def\lb{\mathbb{L}}
\newcommand{\overbar}[1]{\mkern 1.5mu\overline{\mkern-1.5mu#1\mkern-1.5mu}\mkern 1.5mu}
\shorttitle{Stochastic Modeling of Photometric Reverberation Mapping Data}
\shortauthors{Y.\ Zu et al. }
\begin{document}
%%%%%%%%%%%%%%%%%%%%%%%%%%%%%%%%%%%%%%%%%%%%%%%%%%%%%%%
\title{Application of Stochastic
Modeling to Analysis of Photometric Reverberation Mapping Data }
%%%%%%%%%%%%%%%%%%%%%%%%%%%%%%%%%%%%%%%%%%%%%%%%%%%%%%%
\author{
    Ying Zu\altaffilmark{1,2,3},
    C.S.\ Kochanek\altaffilmark{1,2},
    % and
    Szymon Koz\l{}owski\altaffilmark{4},
    and
    B.M.\ Peterson\altaffilmark{1,2}
}
%%%%%%%%%%%%%%%%%%%%%%%%%%%%%%%%%%%%%%%%%%%%%%%%%%%%%%%
\altaffiltext{1}{Department of Astronomy, The Ohio State University, 140
West 18th Avenue, Columbus, OH 43210, USA; yzu@cmu.edu}
\altaffiltext{2}{The Center for Cosmology and Astroparticle Physics, The
Ohio State University, 191 West Woodruff Avenue, Columbus, OH 43210, USA}
\altaffiltext{3}{Department of Physics, Carnegie Mellon University, 5000
Forbes Avenue, Pittsburgh, PA 15213}
\altaffiltext{4}{Warsaw University Observatory, Al. Ujazdowskie 4, 00-478
Warszawa, Poland}
% \altaffiltext{4}{TBD}
%%%%%%%%%%%%%%%%%%%%%%%%%%%%%%%%%%%%%%%%%%%%%%%%%%%%%%%
%
%%%%%%%%%%%%%%%%%%%%%%%%%%%%%%%%%%%%%%%%%%%%%%%%%%%%%%%
%%  Abstract and Keywords
%%%%%%%%%%%%%%%%%%%%%%%%%%%%%%%%%%%%%%%%%%%%%%%%%%%%%%%
\begin{abstract}
  We use both simulated and real quasar light curves to explore
  modeling photometric reverberation-mapping (RM) data as a stochastic
  process. We do this using modifications to our previously developed
  RM method based on modeling quasar variability as a damped random
  walk. We consider the feasibility of one and two-band photometric RM
  and compare the results with those from spectroscopic RM.  We find
  that our method for two-band photometric RM can be competitive with
  spectroscopic RM only for strong (large equivalent width) lines like
  \ha\ and \hb, and that the one-band method is also feasible, but
  requires very high-precision photometry.  We fail to robustly detect
  \ha\ lags in single-band quasar light curves from OGLE--III and IV
  despite the outstanding cadence and time span of the data, on
  account of photometric uncertainties in the range
  $0.02$--$0.04$\,mag. Simulations suggest that success could be
  achieved if the photometric uncertainties were of order 0.01\,mag.
  Single-band RM for all lines and two-band RM for lower equivalent
  width lines are likely only feasible for statistical estimates of
  mean lags for large samples of AGN of similar properties~(e.g.,
  luminosity) rather than for individual quasars. Our approach is
  directly applicable to the time-domain programs within ongoing and
  future wide-field imaging surveys, and could provide robust lag
  measurements for an unprecedented number of systems.
\end{abstract}
%%%%%%%%%%%%%%%%%%%%%%%%%%%%%%%%%%%%%%%%%%%%%%%%%%%%%%%

\section{Introduction}
\label{sec:intro}

Determination of the masses of the supermassive black holes (SMBHs) over cosmic history
is of interest for a number of reasons. Unfortunately, most methods of measuring SMBH masses
rely on high angular resolution and are thus currently feasible only in the local universe. It is possible, however,
to measure the masses of the SMBHs at the centers of active galactic nuclei (AGNs), or quasars, by use of
reverberation mapping \cite[RM;][]{blandford1982, peterson1993,peterson2014}, which substitutes time resolution
for angular resolution. While AGNs constitute only a trace population, they seem to show, for example,
the same $M_\mathrm{SMBH}$--$\sigma_*$
relation \citep[][]{ferrarese2000, gebhardt2000-1,
ferrarese2001, nelson2004, onken2004, dasyra2007, woo2010, graham2011, park2012,grier2013a}
that has been driving studies of galaxy--SMBH co-evolution~\citep{silk1998, fabian1999, king2003, king2005,
treu2004, murray2005, di_matteo2005, di_matteo2008, peng2006, shankar2009, shankar2009-1, merloni2010}. Moreover, the highest redshift quasars \cite[e.g.,][]{mortlock2011,DeRosa2014,carnall2015,jiang2015}
put strong constraints on the SMBH formation and
growth in the early universe \cite[e.g.,][]{volonteri2010,latif2013,latif2015}.

At the present time, published RM studies have for the most part been restricted to
local AGNs \citep[e.g.,][]{peterson2004, denney2010,grier2012-1,bentz2013,barth2015,DeRosa2015}. Standard
RM relies on the spectroscopic monitoring of broad-line fluxes, which are then cross-correlated with a
continuum light curve, either also measured from the spectra, measured independently
photometrically,  or constructed from some combination of the two,
to obtain the ``lag'' or time delay between continuum flux variations and the emission-line
response. The lag is taken to be the light travel time from the central engine to the broad-line
region (BLR). By combining the
lag with a suitable measurement of the emission-line width
$\Delta V$, which is taken to be
an estimate of the virial motion of the BLR gas, one obtains an estimate for the mass
of the SMBH
\begin{equation}
    M_\mathrm{SMBH} = f \frac{c\tau\Delta V^2}{G},
\end{equation}
where $f$ is a dimensionless factor of order unity that depends on the geometry, inclination, and
kinematics of the BLR. The factor $f$ can in principle be determined for individual AGNs by
modeling the BLR \citep{pancoast2011,pancoast2014}.

At the present time, it is common to use the AGN $M_\mathrm{SMBH}\text{--}\sigma_*$ relationship to establish
a mean value for this scaling factor; the most recent published value is
$\langle f \rangle = 4.31 \pm 1.05$ for a line dispersion estimate of $\Delta V$~\citep{grier2013a}.
While use of an average value $\langle f \rangle$ cannot be
expected to yield particularly accurate masses for individual sources,
it is useful for application to large data sets.

RM results have also established an empirical relationship between the AGN luminosity $L$
and the BLR size $R_{\rm BLR}$ of the form
\begin{equation}
\label{eq:RL}
R_{\rm BLR} \propto L^{\alpha},
\end{equation}
where $\alpha \approx 0.5$
\citep{Wandel1999,kaspi2000,kaspi2005,bentz2006,bentz2009,bentz2013}.
This relationship has been independently established through microlensing studies \citep{guerras2013}.
Its particular value is that the AGN luminosity can thus be used as a proxy for measuring
the BLR radius, thus bypassing resource-intensive RM.

It is nevertheless desirable in many cases to measure $R_{\rm BLR}$ and $M_{\rm SMBH}$
directly. However, it is clear that more efficient methods must be found if one wishes to
extend RM to higher-redshift, fainter objects. One possibility that is being actively pursued
is to use multi-object spectrographs on large telescopes to monitor as many as hundreds
of quasars simultaneously \citep{Shen2015,king2015}. With a large-enough telescope and a wide-enough field of view,
the surface density of quasars is high enough to make spectroscopic multiplexing effective.

Another possibility is to make RM measurements with purely photometric data, using either
narrow-band~\citep{haas2011, pozo_nunez2012} or broad-band~\citep{cd12, chelouche2012, edri2012} filters.
The narrow-band approach essentially reduces to the spectroscopic RM problem by subtracting the continuum
contribution to the narrow photometric band containing the broad emission line.

However, the broad-band case is considerably more complicated because
the emission lines and continuum are not easily separable. In this case
the continuum variability has to be carefully modeled or
removed statistically. \cite{cd12} developed a method by looking for an excess cross-correlation signal at
non-zero time lags between two broad-band light curves and applied it to a subset of the Palomar--Green
quasar sample, finding broad agreement with the spectroscopic RM results, albeit with very large
uncertainties. The cross-correlation of two bands, one on and one off the line, or the auto-correlation of
one band on the line, must mathematically have power near both zero lag due to the correlation of the
continuum or line variability with itself and at the emission-line lag due to the cross-correlation between
the line and the continuum. This by no means implies the existence of separate peaks, but only that the
presence of the line emission leads to a broader correlation function. \citeauthor{chelouche2013}~(2013,
hereafter CZ13) proposed a comprehensive approach based on cross-correlation functions~(CCF), which takes
into account the continuum time lag between two broad bands and estimates model uncertainties using Monte
Carlo simulations.  However, discussions of photometric RM to date have not quantitatively addressed the
ability to recover lags accurately as a function of line strength, sampling, and signal-to-noise ratio or made
detailed comparisons to the performance of spectroscopic RM. Moreover, while the work of CZ13 shows
that the existence of a lag can be detected, whether or not it can be
measured with sufficient accuracy that the method becomes competitive with spectroscopic RM remains
dubious. While the problems of photometric RM are both fairly obvious and significant, thorough investigation
of the technique is warranted  since, as pointed out by CZ13, the community will soon be awash in
photometric monitoring data on quasars that we must be ready to exploit fully.

Here we consider a somewhat different approach to extracting lags from photometric data alone.
A series of recent studies~\citep[][]{kelly2009, kozlowski2010, macleod2010, zu2013, andrae2013} have shown
that quasar optical variability is well modeled by the damped random walk (DRW) stochastic process on time
scales from several days to years, although there may be deviations on shorter time
scales~\citep{mushotzky2011, zu2013}.  \citeauthor{zu2011}~(2011; hereafter ZKP11) further adapted
the DRW model to address RM,  where it has several significant advantages over
standard methods.
First, irregularly sampled light curves require some method of interpolation for any inter-comparison, where
the standard approaches use either binning~\citep{edelson1988} or linear interpolation~\citep{gaskell1986}. The ZKP11 approach essentially averages
over all possible continuous light curves that are statistically consistent with the observed data and the DRW
(or other) stochastic process~\citep{rybicki1992}. Second, the means of estimating likelihoods in the various
CCF methods are fairly {\it ad hoc}, while the ZKP11 methods use likelihood functions that can be interpreted
using standard Bayesian or frequentist methods.  Finally, the approach allows generalizations
that can automatically
include calibration uncertainties, temporal trends, data correlations or multiple lines or line velocity bins
in the full likelihood calculation. We have made the analysis software public\footnote{\javelin\
({\tt SPEAR})
\url{http://bitbucket.org/nye17/javelin}} and it is increasingly being used in recent RM
studies~\citep{grier2012-1, grier2012, grier2013b, dietrich2012, zhang2013, shapovalova2013,
li2013,shappee2014}. The spectroscopic RM module in the software has also provided important cross-checks to some
narrow-filter-based photometric RM studies~\citep[e.g.,][]{pozo_nunez2013}.

For the sake of completeness, we also consider the greater challenge of single-band photometric RM:
a reverberating emission line in a photometric band will result in an ``echo'' of the continuum
variations and in principle, if the DRW is indeed a good model for the continuum variability, the
emission-line echo should be identifiable, given sufficient high-quality data.

Here we will apply our modeling
approach to a comparison of spectroscopic and both two and single-band photometric
RM. In particular, we address whether photometric RM is more or less observationally
efficient than spectroscopic RM. We will first summarize our approach in \S\ref{sec:method}.  In
\S\ref{sec:ideal}, we carry out a series of Monte Carlo simulations for all three approaches as a function of
cadence, campaign duration, and line strength relative to the continuum.  In \S\ref{sec:pg}, we examine
two-band photometric RM of PG 0026+129 for which contemporaneous spectroscopic and photometric
light curves are available. In \S{\ref{sec:5548}, we use photometric data on the well-studied
Seyfert galaxy NGC\,5548 as a second case study. In \S\ref{sec:ogle}, we examine single-band photometric
RM using OGLE~(Optical Gravitational Lensing Experiment) quasars behind the Magellanic Clouds.
We summarize our results in \S\ref{sec:conclude} and consider the applicability of these
methods in large-scale surveys such as LSST.

\section{Methodology}
\label{sec:method}

In any given photometric band, quasar variability consists of two components, one from the continuum and the
other from broad emission lines, plus contaminants such as the host-galaxy flux and narrow emission lines that
do not vary on the relevant time scales~\citep[e.g.,][]{peterson2013-1}.  We model the continuum variability
as a Gaussian process~\citep{rasmussen2006, kelly2013, zu2013}
\begin{equation}
    c(t) = \gp\{\overbar{c} , \kappa(t, t')\},
\end{equation}
where the mean function $\overbar{c}$ is constant and $\kappa(t, t')$ is the covariance function between two
epochs.  For the DRW model discussed in \S\ref{sec:intro}, $\kappa(t, t')=\sigma^2\exp(-|t-t'|/\taud)$ where
$\sigma^2$ and $\taud$ are the variance and characteristic time scale of the process. The variability of the
broad emission lines relative to the continuum can be described as
\begin{equation}
    l(t) = \int \Psi(t-t')c(t)\,dt' ,
\end{equation}
where $\Psi(t)$ is the transfer function. In this paper we focus on cases in which there is only one dominant
broad line or multiple broad lines that have similar lags. In particular, following ZKP11, we explicitly model
$\Psi(t)$ as a top hat transfer function centered on time lag $\tau$ with width $w$ and amplitude $A$, so that
\begin{equation}
    \Psi(t)\equiv \Psi(t|\tau, A, w) =  A / w \;\;\mbox{for}\;\; \tau - w/2 \leqslant t < \tau + w/2.
    \label{eqn:tophat}
\end{equation}
Here we consider two classes of monitoring bands: a continuum band~(hereafter referred to as the $\cb$-band)
uncontaminated by lines, and a line band~(hereafter referred to as the $\lb$-band) containing both line and
continuum contributions. The light curves in the $\cb$- and $\lb$-band light curves are $f_{cb}=c(t)+u_{cb}$
and $f_{lb}=\alpha \cdot c(t)+l(t)+u_{lb}$, respectively, where $\alpha$ is the ratio between the continuum
variabilities in the two bands and $u_{cb}$ and $u_{lb}$ represent any contaminating flux from narrow emission
lines and the host galaxy. The key problem for photometric RM methods is whether we can distinguish $l(t)$
from $c(t)$ without measuring $l(t)$ directly as is done in spectroscopic RM. Depending on the available data,
there are two possible approaches to photometric RM.
\begin{itemize}
    \item We could have both $\cb$ and $\lb$ bands. In this case, The
    $\cb$-band light curve provides an independent constraint on the
    structure and statistics of the continuum variability.  Such a model has
    six parameters, $\mathbf{p}\equiv\{\sigma, \taud, \tau, w, A, \alpha\}$,
    where $u_{cb}$ and $u_{lb}$ are nuisance parameters marginalized in the
    analysis.
    \item If we have only the $\lb$ band where $f_{lb}(t) = \alpha \cdot c(t)
    + l(t) + u_{lb}$, then the continuum and line variabilities have to be inferred
    simultaneously. Compared to the two-band model, the number of parameters
    in the one-band model is fewer~($\alpha$ is fixed to be unity so
    that $\mathbf{p}\equiv\{\sigma, \taud, \tau, w, A\}$) but the difficulty
    is substantially increased due to the lack of independent information
    on the continuum variability as compared to the lines.
\end{itemize}
We explore both of these approaches using the statistical framework described by ZKP11. Here we briefly
summarize this approach, and readers should refer to ZKP11 for additional details.

Let $\mathbf{y}$ be a vector comprised of all the light curves, either the combined $\cb$- and $\lb$-band
light curves or the one $\lb$-band light curve. We model $\mathbf{y}$ as
\begin{equation}
\mathbf{y}(t) = \mathbf{s}(t) + \mathbf{n} + L\mathbf{q},
\end{equation}
where $\mathbf{s}(t)$ is the underlying variability signal with zero mean~(e.g., $c(t)-\overbar{c}$ in
$\cb$-band) and covariance matrix $S$,\footnote{The entries of $S_{ij}$ are simply the values of the
covariance function $S_{ij}=S(\dt_{ij})$, so we have used the same symbol for both.} $\mathbf{n}$ is the
measurement error with covariance matrix $N$, and $L$ is a $x\times K$ matrix where $x$ and $K$ are the
number of light curves and total number of data points in $\mathbf{y}$, respectively. In particular, for the
two-band model, $L$ has entries of $(1,0)$ for the $\cb$-band data points, and $(0,1)$ for the $\lb$-band
data points, while for the one-band model $L$ is a vector of all ones. The linear coefficients $\mathbf{q}$
are the light curve means, including contributions from $\overbar{c}$, the mean of $l(t)$, and the host galaxy
light and narrow line flux ($u_{cb}$ and $u_{lb}$).

\begin{figure*}[t]
\epsscale{0.80} \plotone{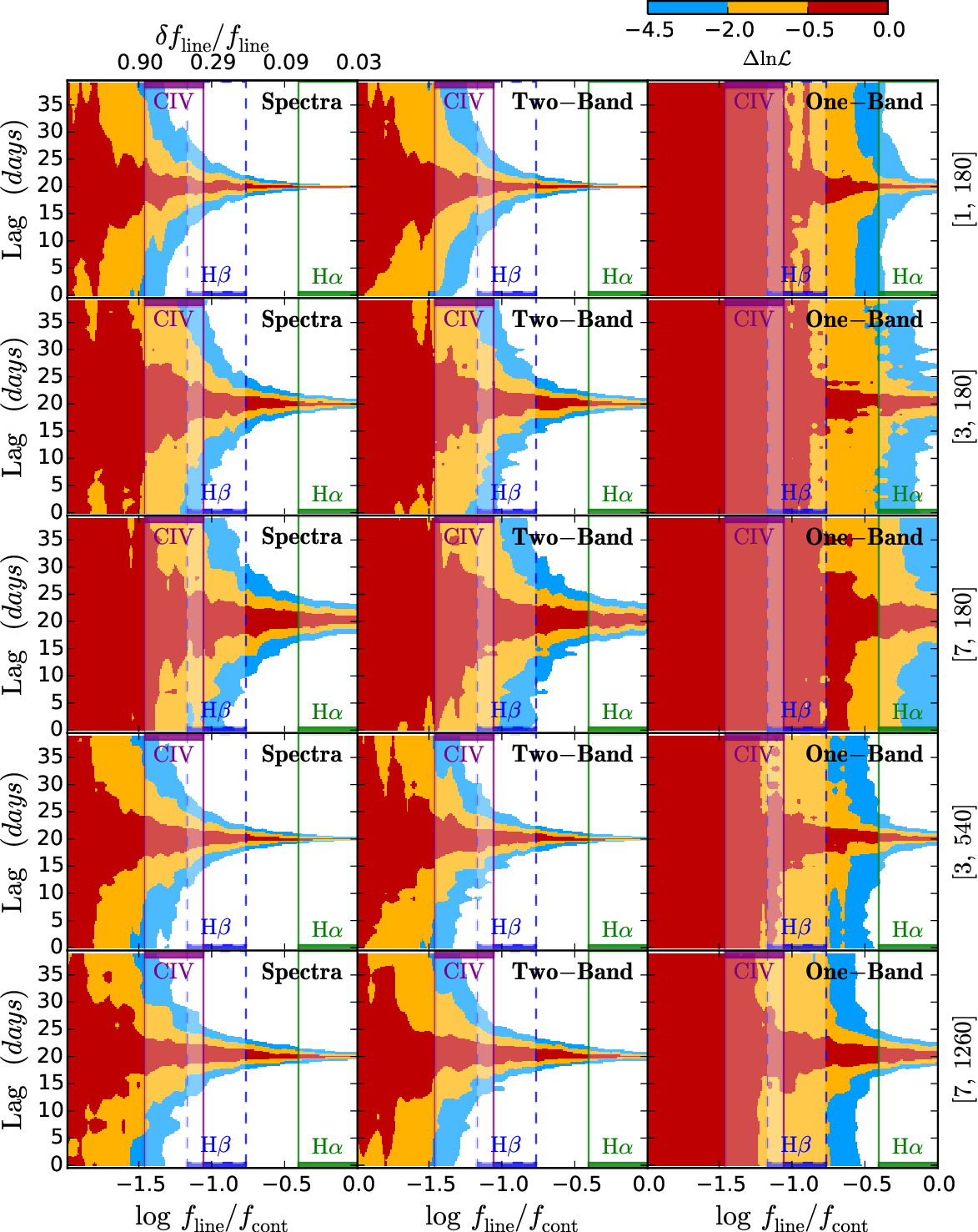} \caption{Significance of lag
detection as function of broad line strength in
    spectroscopic~(left), two-band~(middle) and one-band~(left) RM.
    Panels in the same row assume the same cadence and duration, labeled in the
    format [cadence, duration] on the right of each row. The contours are color-coded by the colorbars
    shown at top, with the red, yellow, and blue contours roughly corresponding
    to the $1\sigma$, $2\sigma$, and $3\sigma$ confidence regions for one
    object, respectively. If one could combine the likelihoods of multiple
    identical objects in each panel, the corresponding confidence regions
    for detecting a {\it mean} lag will shrink considerably. For example,
    after stacking nine objects the red contours will represent
    $3\sigma$ limits instead of $1\sigma$~($\Delta\ln\mathcal{L}$ goes from
    $-0.5$ to $-4.5$). There assume $1\%$ measurement uncertainties for the
    continuum band and the top x-axis for the upper left panel shows the
    corresponding fractional uncertainties in the spectroscopic line flux. See
    text for details.}
\label{fig:ideal_cases}
\end{figure*}

As derived by ZKP11, after marginalizing over $\mathbf{q}$, the likelihood of the model parameters is
\begin{equation}
\mathcal{L}(\mathbf{y}|\mathbf{p}) = |C|^{-1/2}|L^T C^{-1}L|^{-1/2}\exp
\left(-\frac{\mathbf{y}^TC_{\perp}^{-1}\mathbf{y}}{2}\right)
\label{eqn:likelihood}
\end{equation}
where $C=S+N$ is the overall data covariance and
\begin{equation}
C_{\perp}^{-1} = C^{-1}-C^{-1}L(L^TC^{-1}L)^{-1} L^TC^{-1}.
\end{equation}
For light curve prediction, the best estimate for the mean intrinsic variability is
\begin{equation}
   \bfhs =  S C^{-1} (\bfy-L\bfhq),
  \label{eqn:shat}
\end{equation}
where
\begin{equation}
   \bfhq =  (L^T C^{-1} L)^{-1} L^T C^{-1} \bfy
\end{equation}
is the best estimate for the light curve means,
and the expected variance in the estimated variability about the mean is
\begin{equation}
   \langle (\bfs-\bfhs)^2 \rangle = S - S^T C_\perp S.
   \label{eqn:svar}
\end{equation}

\begin{figure*}[t]
\epsscale{1.00}
\plotone{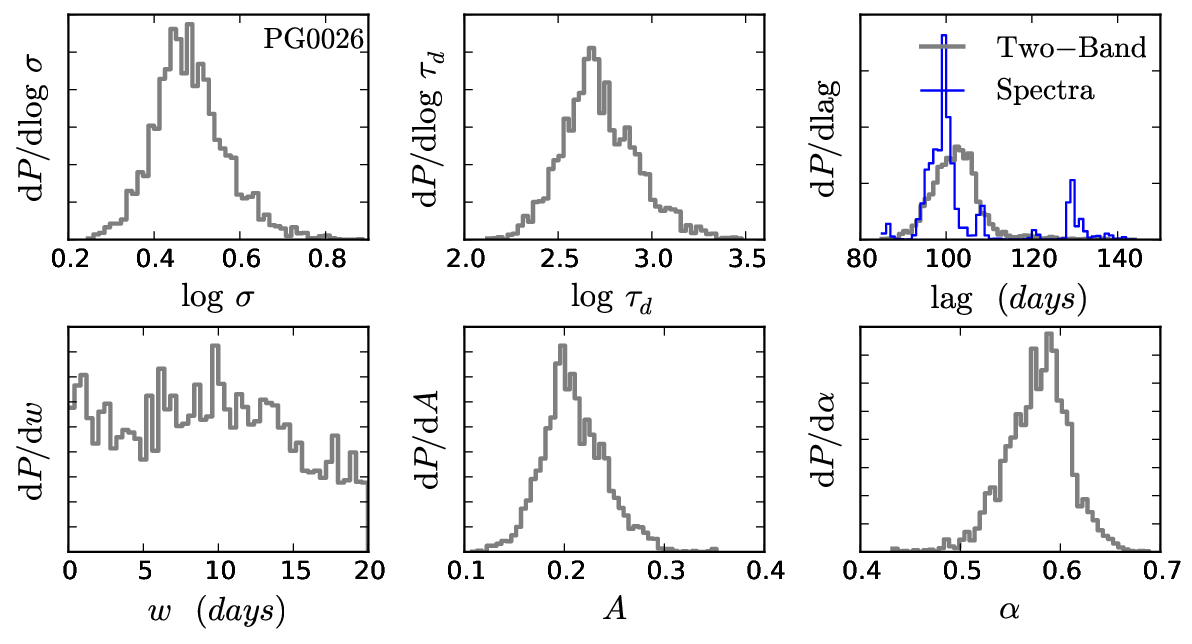}
\caption{Constraints from the photometric RM model using the $B$- and $R$-band light
    curves of PG~0026+129 for the DRW parameters $\sigma$, $\taud$, the lag, the
    kernel width $w$, the transfer function amplitude $A$, and the ratio
    between the continua in the two bands $\alpha$. The thin histogram in the
    top right panel is the estimate of the lag from spectroscopic RM.}
\label{fig:lchain}
\end{figure*}

The only difference between spectroscopic and photometric RM approaches within this framework lies in the
computation of the covariance matrix $S$. For two-band photometric RM, $S_{ij}$ involves three types of
entries, the DRW covariance function $\kappa(t_j-t_i)$, the covariance between $f_{cb}(t)$ and $f_{lb}(t)$
\begin{equation}
    \langle f_{cb}(t_i) f_{lb}(t_j)\rangle =  \alpha \kappa(t_j-t_i) +
    \alpha \langle c(t_i)l(t_j) \rangle ,
    \label{eqn:cifj}
\end{equation}
and the covariance between two $\lb$-band light curves
\begin{align}
    \langle f_{lb}(t_i) f_{lb}(t_j)\rangle = \alpha^2 \kappa(t_j-t_i) +&
    \nonumber\\
    \alpha \langle c(t_i)l(t_j) \rangle +
    \alpha \langle l(t_j)c(t_i) \rangle &+
    \langle l(t_i)l(t_j) \rangle , \label{eqn:fifj}
\end{align}
while for the one-band case, only Equation~\ref{eqn:fifj} is relevant and we can take $\alpha\equiv1$. For
the simple case of top-hat transfer functions, all the terms in Equations~\ref{eqn:cifj} and~\ref{eqn:fifj}
have analytical forms, which can be found in the Appendix of ZKP11.

% 1.0  180.0 180  csknew $19.4\pm2.2$ pmap $20.1\pm2.0$ spmap $20.7\pm11.3$
% 3.0  180.0  60  csknew $20.2\pm5.1$ pmap $20.1\pm6.4$ spmap $18.6\pm10.8$
% 3.0  540.0 180  csknew $20.3\pm2.2$ pmap $19.6\pm2.1$ spmap $20.9\pm 8.0$
% 7.0  180.0  26  csknew $19.2\pm6.7$ pmap $17.2\pm7.8$ spmap $22.8\pm 9.2$
% 7.0 1260.0 180  csknew $20.4\pm1.6$ pmap $19.8\pm1.9$ spmap $21.3\pm 6.8$
\begin{deluxetable}{cccccc}[H]
\tablecaption{\label{table:simu} Lag Estimates for Simulated \hb\ Lines}
\tablehead{\colhead{Cad} & \colhead{Baseline} & \colhead{$N_{p}$} & \colhead{Spec} &
    \colhead{Two-band} & \colhead{One-band} }
\startdata
1 &  180 & 180  & $19.4\pm2.2$ & $20.1\pm2.0$ & $20.7\pm11.3$ \\
3 &  180 &  60  & $20.2\pm5.1$ & $20.1\pm6.4$ & $18.6\pm10.8$ \\
3 &  540 & 180  & $20.3\pm2.2$ & $19.6\pm2.1$ & $20.9\pm 8.0$ \\
7 &  180 &  26  & $19.2\pm6.7$ & $17.2\pm7.8$ & $22.8\pm 9.2$ \\
7 & 1260 & 180  & $20.4\pm1.6$ & $19.8\pm1.9$ & $21.3\pm 6.8$
\enddata
\label{tab:simu}
\tablecomments{Mean lags and their 1-$\sigma$ uncertainties for the simulated \hb\
    lines~($f_{\mathrm{line}}/f_{\mathrm{line}}{=}0.1$) in Figure~\ref{fig:ideal_cases}. ``Cad'' gives the
    observing cadence in days over a period of ``Baseline'' days, producing $N_p$ epochs of observations.
    The input lag is $20$ days.}
\end{deluxetable}

Following ZKP11, we use MCMC methods to estimate the
posterior distributions of model parameters. For the two-band case, we first
constrain $\sigma$ and $\taud$ using the $\cb$-band light curve, and then
apply the $68\%$ confidence limit on each of the two parameters as uncorrelated log-normal
priors to the second step, in which we derive constraints on all the six parameters using the combined $\cb$
and $\lb$-band light curves. As explained in ZKP11, the uncorrelated log-normal priors on $\sigma$ and $\taud$
are necessary to exclude a wrong class of solutions with $\tau_d\to 0$ during the joint fit, and are much more
conservative than the correlated 2D constraints from the $\cb$-band light curve.  For the one-band case,
however, we drop the first step and fit the model to the $\lb$-band light curve directly using uniform priors
on $\log\sigma$ and $\log\taud$. While we do not do so here, a prior on $A/\alpha$, the line strength in the
band, can be added to ``stabilize'' the line contribution. The algorithms for the two photometric RM methods
are implemented in a new update of the \javelin\ software, which is the updated version of \spear\ released
with ZKP11 and is publicly available at \url{http://bitbucket.org/nye17/javelin}.

\section{Application to Simulated Light Curves}
\label{sec:ideal}

Similar to its spectroscopic counterpart, lag detection in photometric RM is very sensitive to the sampling
properties of the light curves, which are mainly characterized by the duration and the cadence of
observations.  Photometric RM further depends critically on the relative strength of the broad lines compared
to the continuum flux within the observational band~(hereafter simply referred to as the ``line strength''),
whereas in spectroscopic RM the separation between line and continuum fluxes is more sensitive to the quality
of spectra than to the line strength.  However, since photometry is more efficient than spectroscopy in
collecting photons~(but only by a factor of ${\sim}2$ for modern spectrographs), given fixed sampling
conditions and exposure time, photometric RM could be competitive with spectroscopic RM.
% Furthermore,
% spectroscopic RM data are subject to time- and wavelength-dependent light-losses that affect differently the
% unresolved nucleus and extended components like host galaxy starlight; This issue is largely eliminated when
% imaging data are used.
Spectroscopic RM data are more difficult to calibrate, due to time- and wavelength-dependent slit-losses and
variable host galaxy contamination. Photometric RM, on the other hand, is (much) more strongly restricted in
the redshift range accessible to an individual observation due to the narrow wavelength coverage of filters as
compared to spectrographs.

To obtain a quantitative understanding of the feasibility of photometric lag detection, we modelled the
traditional spectroscopic and the two photometric RM methods using mock light curves simulated using the
Cholesky decomposition technique described by ZKP11. We considered monitoring baselines of $180$, $540$, and
$1260$ days and cadences of $1$, $3$, and $7$ days as a function of the line strength. The line strength $r$
is characterized by the ratio of line to continuum fluxes in the $\lb$-band,
$r\equiv\overbar{l(t)}/\overbar{c(t)}$. Note that $r$ is a function of both the equivalent width of the line
and the transmission curve of the filter. We considered $20$ log-spaced values of this ratio from $0.01$ to
$1.0$. For each case we generated $50$ random realizations of light curves assuming a typical local Seyfert 1
galaxy at $z{\sim}0$ like NGC~5548 with $\sigma=0.2 \overbar{c}$, $\taud=40$ days, $\tau=20$ days, and $w=2$ days~(cf.,
Figure 10 in ZKP11). We also include the same constant term $u$ in both bands, so that the $\cb$-band light
curve is $f_{cb}(t)=c(t)+u$ and the $\lb$-band light curve is $f_{lb}(t)=c(t)+l(t)+u$. The light curve means
are still independently fit even if the input values are the same. For each photometric light curve, we assume
a
$1\%$ fractional photometric uncertainty in the total band flux~(i.e.,
$\overbar{c}+u$). The impact of any potential deviation from the DRW model on short time scales should be
negligible.

For the simulations, we assume that the continuum and host contributions to the two bands are the same and that
the host contribution is equal to the mean of the quasar continuum. Thus, if $\sigma_c$ is the noise in the
quasar continuum, the noise in the continuum band ($c(t)+u$) is $\sigma_{cb} = 2^{1/2} \sigma_c$, the noise in
the line band is $\sigma_{lb} = (2+r)^{1/2} \sigma_c$, and the noise in the line flux after subtracting the
continuum and the host is $\sigma_l = (4+r)^{1/2}\sigma_c$, assuming similar width for the two bands. We set the fractional error of the continuum band
to $\sigma_{cb}/(u+\bar{c})=\sigma_{cb}/2\bar{c}=2^{-1/2}\sigma_c/\bar{c}=0.01$, which means that the
fractional error in the spectroscopic line flux is $\sigma_l/\overbar{l}=0.02(2+0.5r)^{1/2}/r$.

Figure~\ref{fig:ideal_cases} summarizes the results of these simulations, where we show the {\it average}
likelihood ratio $\ln \mathcal{L}/\mathcal{L}_\mathrm{max}$ expected for a single object as a function of the
input line strength $r$ and the output lag estimate $\tau$ for the spectroscopic~(left column), two-band
photometric~(middle column), and one-band photometric~(right column) methods.  The typical ranges of $r$ for
the \civ\, \hb\, and \ha\ lines and typical broad band filters are indicated by the vertical bands. These
were estimated by convolving the composite quasar spectrum from~\citet[][]{vanden_berk2001} with the
transmission curves of typical broad band photometric systems~(e.g., SDSS or Johnson bands). To avoid clutter,
we do not show the $r$ range for the \mg2\ line. It would lie between the ranges for \civ\ and \hb.

The top three panels in Figure~\ref{fig:ideal_cases} show the forecasts for our fiducial monitoring campaign
--- daily cadence over one observing season~(i.e., six months). The mock light curves used for the second and
third rows have lower sampling rates, with 3 and $7$--day cadences, respectively, while having the same
overall temporal baseline as the fiducial campaign. The mock light curves used for the fourth and fifth rows
have lower sampling rates but longer baselines, maintaining the same number of epochs~($180$) as in the first
test~(the top row). Unsurprisingly, for any given sampling of the light curves~(i.e., comparing panels in the
same row), spectroscopic RM performs better than the two-band photometric RM, and both of them are
significantly better than the one-band photometric RM. Table~\ref{tab:simu} summarizes the result for $r{=}1$~(\hb).
For example, the lag estimates for the fiducial campaign are $19.4\pm2.2$, $20.1\pm2.0$, and $20.7\pm11.3$ days for the
spectroscopic, two-band, and one-band photometric RM cases. For this strong line, the two-band approach
is competitive with spectroscopy, but the one-band approach is not.
% for the fiducial campaign the lag
% estimate for a line with $r=0.1$ is $19.4\pm2.2$ days and $20.1\pm2.0$ days~($1\sigma$ level) in the spectroscopic
% and two-band photometric RM, but degrades drastically in the one-band tests to $20.7\pm11.3$ days~(see
% Table~\ref{tab:simu} for the comparison in other cases).
Within each method~(i.e., comparing panels in the same column), the lag detection efficiency is very sensitive to the
cadence for any fixed baseline, but largely due to the decrease in the number of data points for longer
cadences --- the difference among the first, the fourth, and the bottom panels in each column, where the total
number of epochs is fixed to $180$, is much smaller than that among the top three panels where the number of
epochs is varied. In particular, in the middle column for the two-band test, the fiducial and the two
long-baseline tests~(bottom two panels) yield very similar lags of $20\pm2$ days for lines
with $r=0.1$, while the two short-baseline tests~(second and third panels) find $20\pm6$ days or worse,
respectively. However, we expect the uncertainties in the lag estimates to rise rapidly even for a fixed
number of epochs as the sampling rate decreases once the cadence is larger than $20$ days.

The uncertainties in the lags are also highly sensitive to the magnitude of the line contribution in the
photometric band. As can be seen in Figure~\ref{fig:ideal_cases}, this means that for campaigns with $180$
epochs, \ha\ lags are almost always measurable with high significance using photometric RM methods, \hb\
lags are only marginally measurable, and the \mg2\ and \civ\ lags are never measurable for light curves with
similar sampling properties and physical parameters.  Single-band RM is very challenging even for \ha\ and
relatively ``narrow'' broad-band filters such as the SDSS system~\citep{fukugita1996}.

\section{Case Study I: Two-Band Photometric RM of PG\,0026+129}
\label{sec:pg}
\begin{figure*}[t]
\epsscale{1.00} \plotone{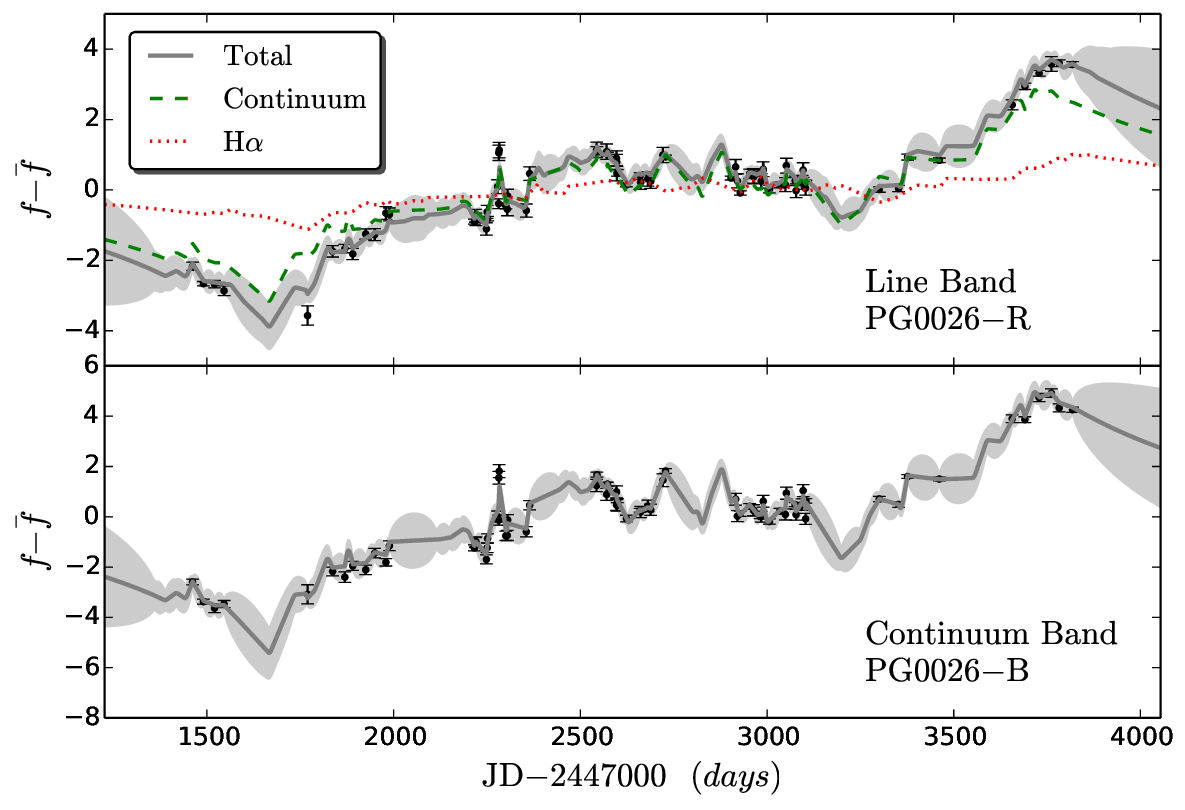} \caption{Comparison between the data and
the mean of the predicted light curves for PG~0026+129, using the best-fitting
parameters from Figure~\ref{fig:lchain}. The lower panel shows the
average of the DRW light curves matching the continuum light curve for the
best-fit parameters and the rms scatter of these light curves. The upper
panel shows the resulting fits to the line band, as well as the decomposition
into the \ha\ line and continuum contributions.} \label{fig:pred}
\end{figure*}

The best chance of obtaining robust photometric lags is to focus on quasars whose \ha\ line lies within one
passband and the continuum is cleanly observed in another passband. The Palomar--Green~(PG) quasar light
curves from \citet{giveon1999} consist of two-band~($B$ and $R$) light curves of $42$ quasars from the PG
sample, with typical cadences of $\sim20$ or $40$ days over a seven-year time span and an average photometric
uncertainty of $0.017$ mag. The sample has a redshift range of $0.1 \lesssim z \lesssim0.3$, making the $B$ and
$R$-band light curves suitable for detecting \ha\ lags using two-band photometric RM (with $B$ being the
$\cb$-band and $R$ being the $\lb$-band). Many of these targets were also contemporaneously monitored in order
to make spectroscopic RM measurements~\citep{kaspi2000}. For these objects, we can assess the performance of
photometric RM by comparing the photometric lag constraints to the corresponding results using spectroscopic
RM. Therefore, we focus on seven objects --- PG\,0026+129, PG\,0052+251, PG\.0804+761,
PG\,0844+349, PG\,1613+658, PG\,1617+175, and PG\,2130+099 ---
that are both photometrically better sampled~(i.e.,
20-day cadence) and have spectroscopic \ha\ light curves.
\cite{cd12} also made photometric RM measurements for several of these systems but with such large
uncertainties~($\sim100$ to $200$ days) that the significance of any comparison is very small.

In each case, we model the continuum-band light curve as
$f_{cb}(t)=c(t)$ and the line-band light curves as $f_{lb}(t)=\alpha\cdot c(t) + \Psi(t) \ast c(t)$ where
$\Psi(t)$ is the top-hat transfer function centered on lag $\tau$ with width $w$ and amplitude
$A$~(Equation~\ref{eqn:tophat}), and $\alpha$ scales the continua between the two bands.
For better consistency in
the modeling of continuum variability, we re-calculated the spectroscopic lags with \javelin\ as described
by \cite{zu2011} using the original \cite{kaspi2000} light curves. The two methods are largely consistent with
each other, with lag differences smaller than twice the average temporal sampling~(i.e., $2\times20$ days) in
all cases except PG\,1613+658. The cause for the discrepancy in PG\,1613+658 is unclear,
as the observed $R$-band light
curve simply cannot be matched by the prediction from \javelin\ using the spectroscopic \ha\ lag.

We applied our two-band photometric RM analyses to all seven of these quasars. Unfortunately, all
of the cases proved to be somewhat problematic, usually because of the low amplitude of \ha\ variability,
but for other reasons as well (e.g., the original lags for PG 2130+099 were badly misidentified; see
\citealt{grier2008,grier2012}).

%Two of these (PG~0026, PG~0052) are at redshifts beyond $0.12$ when the Cousins $R$-band receives
%under $10\%$ of the flux from H$\alpha$. However, as pointed out by CD12, the $R$-band is generally a line-rich
%band with potentially important contributions from the Iron emission lines which reverberate by similar lags
%as the H$\alpha$.  Therefore, we elect to keep them in the sample despite the lines that are responsible for
%the lags might be different from others that are dominated by the H$\alpha$ flux.

Only PG\,0026+129 yielded a plausible result.
Figure~\ref{fig:lchain} shows the posterior probability distributions for the six parameters in
the two-band photometric RM model for PG~0026+129. The distribution of the
top-hat width $w$ is mostly flat within $20$ days, which is the sampling interval of the light curves. The thin
blue histogram in the top right panel shows the posterior probability distribution of the \ha\ lag inferred
from spectroscopic RM. The lag constraints from the two RM methods agree with each other well, indicating that
the photometric RM approach is capable of separating the \ha\ signals from the continuum variability while
recovering the correct lag. The ratio of $A$ to $\alpha$ indicates that the line variability is roughly $1/3$
of the continuum variability within the $R$-band, also consistent with what we expect for \ha\ based on the
spectral template from~\citet{vanden_berk2001}.

One virtue of \javelin\ is that it produces an explicit model for the mean and dispersion of the light curves
constrained by the data given the best-fitting parameters, as shown in Figure~\ref{fig:pred}. In each panel
the observed light curves are shown by the data points with errorbars, while the solid line and the error
``snake'' are the estimated mean of light curves consistent with the data and their variance,
respectively~(Equations \ref{eqn:shat} and \ref{eqn:svar}).  We also show the decomposition of the model for
the $R$-band light curve into the line and continuum contributions, as shown in the top panel, where the
dashed and dotted curves indicate the expected fluxes contributed by the continuum and \ha\ line,
respectively.  It shows unambiguously that the difference between the $B$- and $R$-band light curves can be
well described by a weaker but lagged version of the $B$-band light curve representing the \ha\
light curve.

\section{Case Study II: Photometric Observations of NGC 5548}
\label{sec:5548}

As previously noted, CZ13 proposed a modified CCF-based photometric RM method for simultaneously
estimating the intra-band and line lags. The CZ13 method describes the $\lb$-band light curve as the sum of
two scaled and lagged versions of the $\cb$-band light curve with a non-negligible lag between the continua in
the two optical bands and a delta function for the transfer function of the line.   CZ13 searched through the
3-parameter space~(intra-continuum band lag, line lag, and the line-to-continuum flux ratio) for the model
which maximizes the correlation coefficient $R$ between the observed and predicted $\lb$-band light curves.
Aside from assuming different models for the $\lb$-band light curve, the two key differences between the CZ13
method and \javelin\ are:
\begin{itemize}
\item In order to predict the $\cb$-band light curve values at unobserved epochs, CZ13
employs linear interpolation rather than interpolating in a manner consistent  with the underlying process.
\item To characterize the uncertainties in their parameter estimates, CZ13 generate a sample of mock light
curves as the sum of the data light curves and Gaussian random deviates with the estimated uncertainties as
the dispersions, and then infer the parameters for each mock data set to compute the error distributions.
\javelin\ employs a MCMC Bayesian approach that is self-consistent within the underlying statistical
framework.
\end{itemize}

\begin{figure*}[t]
\epsscale{1.00}
\plotone{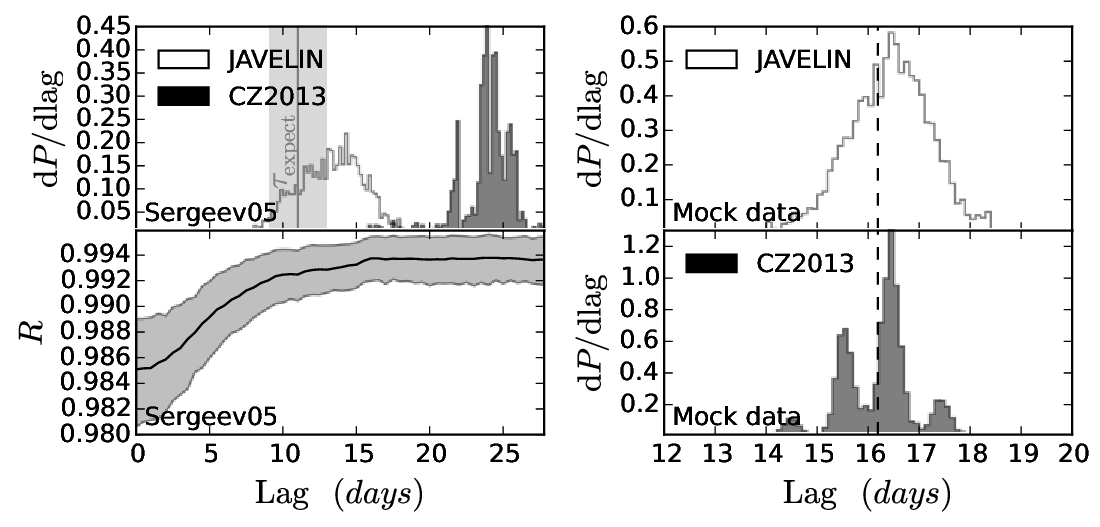}
\caption{Comparison between the \javelin\ and the CCF-based method of CZ13, using
    the NGC~5548 light curves from Sergeev et al. 2005~(left) and mock light
    curves~(right). The vertical band in the upper left panel indicates the
    expected lag given the luminosity state of NGC~5548 at the time of
    observation. See text for details.}
\label{fig:duel}
\end{figure*}

To compare the performance of the CZ13 method and \javelin\, we applied both methods to the two-band~($V$
and $R$) light curves of NGC~5548 from~\citet{sergeev2005} that CZ13 used as their principal example.
While there is no \ha\ light curve for this period, NGC 5548 is sufficiently well-characterized that
the \ha\ lag can be estimated from the AGN luminosity.
NGC~5548 was at a near-historic low-luminosity state at the time of the \cite{sergeev2005}
observations~(2001--2002; see Figure 1 in~\citet{peterson2013-1} for an NGC~5548 continuum light curve
over the past two decades).
The LAMP spectroscopic RM campaign in 2008, when the continuum was
at a similarly low level, yielded lags for  \hb\  and \ha\ of $4.25^{+0.88}_{-1.33}$ days and
$11.02^{+1.27}_{-1.15}$ days, respectively~\citep{bentz2010}. Thus, we expect a similar \ha\ lag at the time of the
\cite{sergeev2005} observations.

The results of our analysis of the NGC 5548 photometry
are shown in the left panels of Figure~\ref{fig:duel}, where the top panel compares the two inferred
lag distributions and the bottom panel shows CZ13's correlation coefficient $R$ as a function of lag using the
best-fitting CZ13 model found at fixed lag.
\javelin\ yields a lag of $\sim14$ days for \ha, in excellent agreement with the luminosity-based
prediction. The CZ13 method yields a lag $\sim24$ days, in poor agreement with both the
predicted value and the \javelin\ result. We note, however, that
our result using the CZ13 method
agrees with the estimate in Figure 7 of CZ13, showing peaks at $> 20$ days, but lag estimates quoted in their
Table 1 are smaller and in better agreement with both the predicted lag and the \javelin\ measurement.
The discrepancy between the CZ13 and \javelin\ estimates of the \ha\
lags can be largely attributed to the different interpolation schemes, modulo the difference in
assumed transfer functions.

The significance of the CZ13 lag detection is illustrated by the $R$ curve in
the bottom left panel of Figure \ref{fig:duel}, with the gray band indicating the $68\%$ uncertainty range in
$R$ derived by bootstrapping the light curve data following their procedures.  Since the line-band flux is
always dominated by the continuum, $R$ is strong even at zero lag~($\overbar{R}_0=0.985$), and gradually
increases until hitting a plateau at $\sim10$ days. Therefore, although the lag distribution derived by CZ13
is longer than $20$ days, the statistical significance of the long lags is barely higher than for
10--20-day lags.

Apart from having a shift in the estimated central lags, the two lag distributions in the top left panel of
Figure~\ref{fig:duel} also have different shapes, with a continuous, quasi-normal distribution for \javelin\
and a discrete array of sharp peaks for the CZ13 method~(it is unclear whether the lag distribution derived by
CZ13 has this spiky feature due to their large temporal bins). To investigate the origin of these differences,
we simulated $5000$ sets of two-band photometric light curves that have the same sampling and error properties
as the NGC~5548 light curves using parameters of $\sigma=0.1\overbar{f_{B}}$, $\taud=566.2$ days, $\tau=16.2$
days, $w=2.0$ days, $A=0.5$, and $\alpha=0.64$. To eliminate any discrepancies caused by assuming a different
transfer function width $w$ and intra-band lag, we choose not to use the best-fitting parameters from the left
panel, which prefers a larger $w$ and a shorter $\tau$, and we impose zero lag between the continua in the two
broad bands. For each set of the mock light curves, we compute lag distributions from the $5000$ experiments
for each of the two methods. The results are shown in the two right panels of Figure~\ref{fig:duel}, where the
two lag distributions recover the input 16.2-day lag~(vertical dashed line) but the shape difference persists.
As expected, the lag distribution derived from \javelin\ in the top right panel is continuous and resembles
the distribution we see in the top left panel for the data.  For the CZ13 test, a close examination of the
pattern in the bottom right panel reveals that the discrete peaks occur half-way between integer-day lags
while the distribution is heavily suppressed at integer day lags. This peculiar pattern is a numerical
artifact caused by the use of linear interpolation and a $w=0$ transfer function~(i.e., a $\delta$ function)
in the CZ13 method.  The Sergeev et al. (2005) observations were obtained on a nightly cadence and we
reproduced that cadence for the mock light curves, so no interpolation is required at integer lags to
calculate the correlation coefficient $R$.  At half-integer day lags, the linear interpolation acts like a
transfer function with $w=1$ day instead of $w=0$, smoothing the light curve, and producing a better
correlation by~(essentially) reducing the fluctuations due to noise. This problem becomes worse when the
quasar variability has a short characteristic time scale compared to the sampling cadence and linear
interpolation becomes an increasingly poor approximation. We think this problem is intrinsic to the method,
but the binning of the lag probability distribution in CZ13~(their Fig.\ 7) does not allow us to cross check
this issue against their results. We note, however, the instability of measuring short delays ($\sim 1$ day)
between two broad-band continua (or equivalently the lack of robustness in small inter-band delay
uncertainties), is a well-known problem in determining gravitational lens time delays~\citep[e.g.,
see][]{tewes2013} even for light curves far superior to those used here.

\section{Case Study III: Single-Band OGLE Quasar Light Curves}
\label{sec:ogle}

\begin{figure*}[t] \epsscale{1.00} \plotone{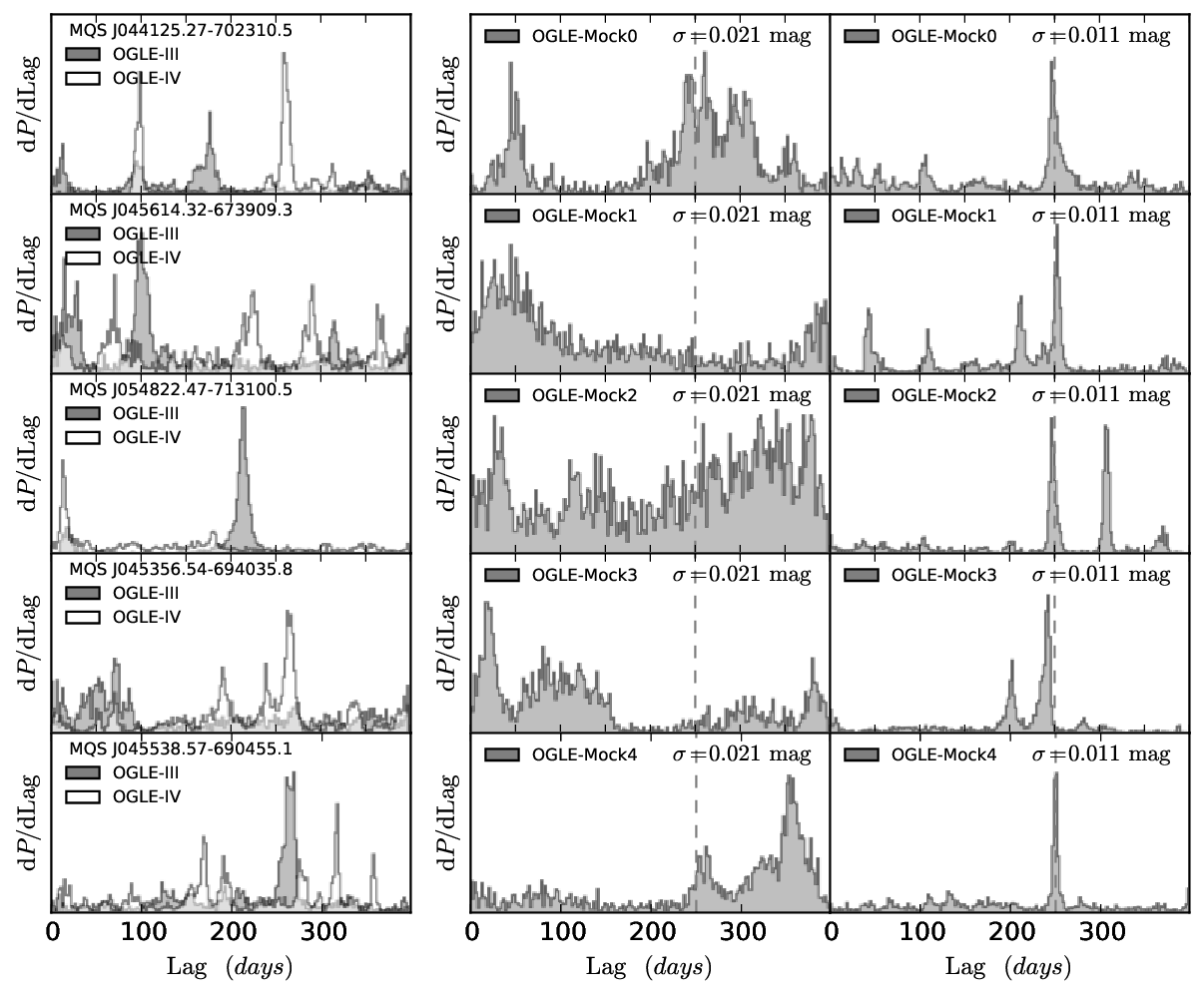} \caption{Lag
        estimates from one-band photometric RM, using either the OGLE--III or
        OGLE--IV light curves~(left), mock light curves using the original
        photometric uncertainties~(middle) and half these photometric
        uncertainties~(right). In the panels using mock data, the input lag
        is indicated by the vertical dashed line.} \label{fig:oglechains}
\end{figure*}

The simulations in \S\ref{sec:ideal} demonstrate that while it is possible to measure \ha\ lags using
single-band photometric RM, the light curves have to be densely sampled with small photometric uncertainties over a
long baseline. Currently one of the best data sets of single-band light curves is from the OGLE
experiment\footnote{There are still better, regular light curves of small numbers of
    AGNs~\citep[e.g.,][]{mushotzky2011}, but the broad \textit{Kepler} filter bandpass makes it poorly suited
to this problem.}, where
quasars behind the Small and the Large Magellanic Clouds~(SMC and LMC) have been monitored with a $\sim 2$ day
cadence for over 8 yrs in the $I$-band by OGLE--III~\citep{udalski2008} and for $\sim 2.5$ years by
OGLE--IV~\citep{soszynski2012, kozlowski2013-1}. Most of these quasars were identified by~\cite{kozlowski2011,
kozlowski2012, kozlowski2013} in part from candidates variability-selected using the DRW
model~\citep{kozlowski2010}.  As a test application of the one-band photometric RM method, we selected $34$
quasars that have prominent, broad \ha\ emission lines that lie in the $I$-band, relatively strong
variability, and at least 400 epochs in OGLE--III. Each quasar also has
 a shorter light curve from OGLE--IV~($\sim350$ epochs). We do not combine
the two OGLE campaigns for the same object, but derive two lags separately as a means of checking the results.
The typical photometric uncertainties of the light curves are $\sim 0.02\text{--}0.04$ mag.

The left column of Figure~\ref{fig:oglechains} shows the lag distributions derived by applying the one-band
photometric RM to the OGLE--III and IV light curves for five random quasars in the sample. In each panel, dark
and light histograms on the left show the constraints from the OGLE--III and IV light curves, respectively.
The quasars are expected to have \ha\ lags of roughly $200\text{--}300$ days based on their optical
luminosity~(i.e., estimated by assuming a typical quasar spectrum normalized by the $I$-band absolute
magnitude using the scaling relations in~\citeauthor{macleod2010} 2010). The five objects all have peaks in
their lag distributions between 200--300 days, however, none of them show consistent lag constraints
between the two OGLE campaigns. The rest of the quasars in the sample all exhibit similar inconsistencies,
indicating that the OGLE light curves collected to date are still insufficient for doing one-band photometric
RM, despite the long baseline and high cadence.

\begin{figure*}[t]
\epsscale{1.00} \plotone{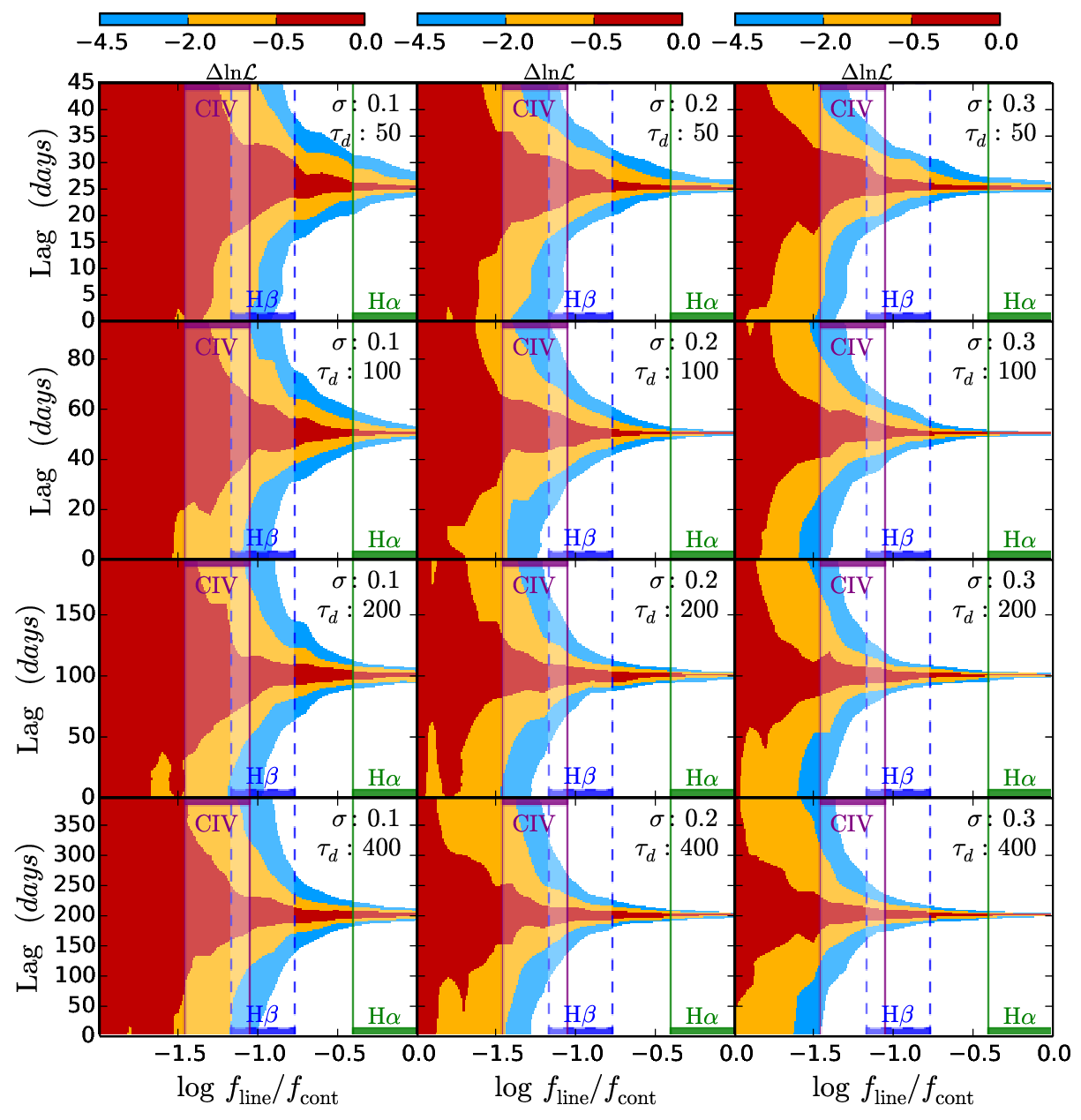} \caption{As in Figure~\ref{fig:ideal_cases},
but for two-band RM using simulated LSST quasar light curves. The variability
parameters $\sigma$ and $\taud$ are
    listed on the top right of each panel, in units of $\mathrm{mag}$
    and $\mathrm{days}$, respectively, and we set the input lag to be
    $0.5\taud$. All the simulated light curves have $200$ epochs, sampled
    on a $14$-day cadence over $10$ years. The photometric uncertainties in the
    continuum band are again $1\%$.
    }
\label{fig:lsst}
\end{figure*}

The non-detection is likely caused by the relatively large photometric uncertainties compared to the line
variability signal. To test the feasibility of one-band photometric RM given OGLE's sampling rates and
photometric errors, we generated two sets of mock light curves that have exactly the same sampling epochs as
object {\tt MQS} {\tt J044125.27-702310.5}~(top left panel of Figure~\ref{fig:oglechains}) using
$\sigma=0.1\overbar{f_{I}}$, $\taud=350$ days, $\tau=250$ days, $w=9$ days, and $A=0.3$. In one case, we used
the original average photometric error of $0.021$ mag and in the other we used half this average error~(i.e.,
$0.011$ mag). We then repeated the analyses for both sets of mock light curves with the results shown in the
middle and right columns of Figure~\ref{fig:oglechains}, where the input lag is indicated by the vertical
dashed line in each panel.  The probability distributions in the middle column
derived from the mock light curves
look less ``spiky'' than those derived from the data light curves in the left column, possibly because
our mock light curves ignore the flux contributions from broad lines other than \ha.  Nonetheless, the
one-band RM method fails to detect the \ha\ lags for the mock light curves with the same photometric errors
as the OGLE data, but unambiguously recovers the input lags in all five cases after the photometric uncertainties
are reduced by half, as shown in the right column of Figure \ref{fig:oglechains}.
Therefore, the key problem for one-band photometric RM is that it needs very
high-precision photometry compared to what is usually obtained for typical sources in large-scale variability
surveys.

\section{Conclusions}
\label{sec:conclude}

We have developed a stochastic modeling approach to analysis of
photometric RM  using the statistical framework introduced by ZKP11, assuming the
continuum variability is described by a DRW model and the line transfer function is a top hat. The approach can
be applied either for two-band photometric RM, where there is a line band and an independent continuum band,
or to one-band photometric RM, where there is only a line band.

By applying the spectroscopic and the two- and one-band photometric RM methods to a suite of simulated light
curves, we find that two-band photometric RM can be competitive with spectroscopic RM only for strong (large
equivalent width) lines like \ha\ and \hb\ in terms of lag detection efficiency for fixed sampling conditions.
The one-band method, however, requires light curves of much higher photometric quality than is generally
achieved and is thus very challenging for any line.  For all three methods we also find that when the average
sampling interval is smaller than the lag, the lag detection significance is most sensitive to the total
exposure time accumulated over all the monitoring epochs and almost independent of cadence.

Application to test cases show that the photometric and spectroscopic lag estimates are broadly
consistent with each other and have comparable lag uncertainties. We also used the one-band photometric RM
approach to analyze a sample of variable OGLE quasars with strong \ha\ emission. The lag estimates from
separately analyzing the OGLE--III and OGLE--IV light curves generally do not agree.  One-band RM fails
even for
the quasars with some of the best-existing long-term light curves. Simulations show that the problem is that the
0.02--0.04 mag photometric uncertainties of the OGLE quasar light curves are simply too large, but
that the one-band method may succeed if the uncertainties were reduced to $\sim0.01$ mag.
Single-band RM is likely most promising for computing average lags for samples of quasars with similar
physical properties~(e.g., luminosity) rather than for individual objects.

In our simulations, we find that for the same observing cadence, two-band photometric RM requires measurement
uncertainties $\xi=0.85$ times smaller than spectroscopic RM to measure a lag with the same accuracy. A common
argument for photometric RM is that it requires far less telescope time, so it is interesting to examine this
claim quantitatively.  Let $t_0$ be the time required to reach a given signal-to-noise level with a
spectrograph that has the efficiency of an imager. Spectrographs are less efficient than
imagers\footnote{Aside from less photon-collecting efficiency, further disadvantages of spectrographs include
    slit-loss and the high S/N required for the measurements of narrow [OIII] lines used for internal
    photometric calibrations.}, and the
difference for modern low-resolution spectrographs is a factor of $\epsilon^{-1} \sim 2$. Thus, the actual
integration time required for spectroscopic RM is $t_S = \epsilon^{-1} t_0$. Two-band photometric RM requires
two images and requires exposure times of $\xi^{-2}$ longer because higher signal-to-noise ratio data are
required to achieve the same lag precision, leading to a total integration time of $t_I=2\xi^{-2}t_0$.  Thus,
the ratio of the required integration times is $t_I/t_S = 2 \epsilon \xi^{-2} \approx 1.48$. Thus, the
advantage of two-band photometric RM over spectroscopy is by no means obvious. Adding target acquisition times
may ultimately favor imaging because of more demanding telescope pointing requirements for spectroscopy, but
only because smaller telescopes have not adequately invested in minimizing such overheads. Other
considerations are that spectra are more difficult to calibrate due to variable slit-losses and host
contamination, while photometric filters strongly restrict the accessible redshift range per observation
compared to spectroscopy.

The conclusion that there is no particular benefit to photometric RM over spectroscopic RM for a single object
contradicts a growing ``conventional wisdom''.  The issue is that conventional wisdom is based on the
integration times that spectroscopic RM campaigns actually use compared to imaging integration times rather
than the integration time they could get away with if all that is desired is an average lag.  For the latter
purpose, spectroscopic RM campaigns are grossly over-integrating, in large part because their present day
goals are focused on measuring lags as a function of velocity within the line~\citep{denney2009, bentz2010,
grier2013b}.  The real potential gain for photometric RM is that wide-field imaging may allow the measurement
of lags for many objects in parallel, but even there, the advantage does not trivially lie with imaging --- as
noted earlier, pilot RM programs  \citep{Shen2015,king2015} are already being pursued with wide-field
spectrographs like SDSS/BOSS~\citep{dawson2013} or AAT/AAOmega~\citep{sharp2006} that are better matched to
the surface density of quasars on the sky than are existing wide-field imagers.

The problem with using wide-field spectrographs for RM is largely sociological --- in any form,
RM requires a large commitment of telescope time, regardless of the size of the telescope, and
competition for large, wide-field telescopes is fierce. In contrast, there are imaging telescopes dedicated
to photometric surveys, including monitoring programs such as
DES~\citep{the_dark_energy_survey_collaboration2005}, Pan--STARRS~\citep{kaiser2002}, and ultimately
LSST~\citep{lsst_science_collaboration2009}. LSST will monitor millions of quasars in six filters over a
ten-year baseline, with $200$ visits per filter each year. Figure~\ref{fig:lsst} shows the expected lag
detection significance
for two-band RM using simulated LSST light curves~($14$--day cadence with $1\%$
photometric uncertainties) using quasar variability parameters of $\sigma=0.1$, $0.2$, and $0.3$ mag and DRW
time scales of $\taud=50$, $100$, $200$, and $400$ days. In each panel, we set the input lag to be $0.5\taud$,
which can always be robustly measured for \ha\ using LSST. The \hb\ lags can only be measured
within $10\%$ for light curves with $\sigma>0.2$ mag and lag $>100$ days, while
the \civ\ signals are still too weak to detect. However, since the \javelin\ method is based on
likelihoods, unlike the CCF methods proposed by~\cite{fine2012, fine2013}, it is straightforward to
multiply the likelihoods of individual quasars to calculate an average lag for quasars of similar
luminosity and redshift, which according to the luminosity--radius relation (Equation~\ref{eq:RL})
should share similar emission line lags. For example, we can multiply the \hb\ or \civ\ lag likelihood
functions of LSST quasars in the same bin of redshift and luminosity and calculate an average lag for that
bin. In some sense, this is what we see in Figures~\ref{fig:ideal_cases} and \ref{fig:lsst}, which show
the average likelihoods expected for a single quasar --- the combined likelihood for $9$ similar quasars
would be the same distribution after multiplying the likelihood ratios by $9$ with consequent narrowing
of the confidence regions, so that the red contours in Figure~\ref{fig:ideal_cases} and~\ref{fig:lsst}
would approximately represent $3\sigma$ confidence regions~($\Delta\ln\mathcal{L}=-4.5$) instead of
$1\sigma$~($\Delta\ln\mathcal{L}=-0.5$). This composite photometric RM method will be particularly
useful for ``piggybacking'' on surveys that produce a large number of under-sampled quasar light curves.
Therefore, provided the systematic errors in photometry can be controlled to the sub-$0.01$ mag level,
we expect that the number of RM systems will grow dramatically with the incoming high-quality photometric
quasar light curves in the near future.

\section*{Acknowledgements} We thank the second referee for helpful comments that have greatly improved the
manuscript. We thank Kelly Denney, Stephan Frank, Catherine Grier, Matthias Dietrich, Jan
Skowron, and Richard Pogge for many helpful discussions. YZ is supported by an OSU Distinguished University
Fellowship, CSK is supported by NSF Grant AST-1009756, SK is supported from (FP7/2007-2013)/ERC grant
agreement no. 246678 awarded by the European Research Council under the European Community's Seventh Framework
Programme to the OGLE project, and BMP is supported by NSF grant AST-1008882.

%%%%%%%%%%%%%%%%%%%%%%%%%%%%%%%%%%%%%%%%%%%%%%%%%%%%%%%
%  Bibliography
%%%%%%%%%%%%%%%%%%%%%%%%%%%%%%%%%%%%%%%%%%%%%%%%%%%%%%%
\bibliographystyle{apj}
% \bibliography{photoreverb}

%
%%%%%%%%%%%%%%%%%%%%%%%%%%%%%%%%%%%%%%%%%%%%%%%%%%%%%%%
\end{document}